\documentclass[amssymb,amsmath,aps,showpacs,floatfix,nofootinbib,showpacs,12pt,prd]{revtex4}

\usepackage{graphicx}
\usepackage{latexsym}
\usepackage{color}

\def\be{\begin{equation}}
\def\ee{\end{equation}}
\def\bea{\begin{eqnarray}}
\def\eea{\end{eqnarray}}

\begin{document}

\title{A Markov Chain Monte Carlo Study on Dark Matter Property Related
to the Cosmic e$^{\pm}$ Excesses}

\author{Jie Liu${}^{a}$}
\author{Qiang Yuan${}^{b}$}
\author{Xiaojun Bi${}^{b}$}
\author{Hong Li${}^{a,c}$}
\author{Xinmin Zhang${}^{a,c}$}

\affiliation{${}^a$Theoretical Physics Division, Institute of High
Energy Physics, Chinese Academy of Science, P.O.Box 918-4, Beijing
100049, P.R.China}

\affiliation{${}^b$Key Laboratory of Particle Astrophysics,
Institute of High Energy Physics, Chinese Academy of Science,
P.O.Box 918-3, Beijing 100049, P.R.China}

\affiliation{${}^c$Theoretical Physics Center for Science Facilities
(TPCSF), Chinese Academy of Science, P.R.China}


\begin{abstract}

In this paper we develop a Markov Chain Monte Carlo code to study the 
dark matter properties in frameworks to interpret the recent 
observations of cosmic ray electron/positron excesses. We assume that 
the dark matter particles couple dominantly to leptons and consider 
two cases, annihilating or decaying into lepton pairs, respectively. 
The constraint on the central density profile from H.E.S.S. observation
of diffuse $\gamma$-rays around the Galactic center is also included 
in the Markov Chain Monte Carlo code self-consistently. In the numerical 
study, we have considered two cases of the background: fixed $e^+e^-$ 
backgrond and the relaxed one. Two data sets of electrons/positrons, i.e. 
PAMELA+ATIC (Data set I) and PAMELA+Fermi-LAT+H.E.S.S. (Data set II), 
are fitted independently, considering the current inconsistence between 
the observational data. We find that for the Data set I, dark matter with 
$m_{\chi}\approx0.70$ TeV for annihilation (or $1.4$ TeV for decay) and 
a non-negligible branching ratio to $e^+e^-$ channel is favored;
while for the Data set II, $m_{\chi}\approx 2.2$ TeV for annihilation 
(or $4.5$ TeV for decay) and the combination of $\mu^+\mu^-$ and 
$\tau^+\tau^-$ final states can best fit the data. We also show that
the background of electrons and positrons actually will significantly
affect the branching ratios. The H.E.S.S. observation of $\gamma$-rays 
in Galactic center ridge puts a strong 
constraint on the central density profile of the dark matter halo for 
the annihilation dark matter scenario. In this case the NFW profile 
which is regarded as the typical predication from the cold dark 
matter scenario, is excluded with a high significance ($>3\sigma$). 
For the decaying dark matter scenario, the constraint is much weaker.

\end{abstract}


\maketitle


\section{Introduction}
\label{Int}

The recent reported results on the abnormal excesses of cosmic ray
(CR) positron fraction by PAMELA \cite{Adriani:2008zr} and the
spectra of electrons\footnote{When we mention about experimental
data of electron spectra, it actually means the total spectra of
electrons and positrons.} by ATIC \cite{Chang:2008zzr}, PPB-BETS
\cite{Torii:2008}, H.E.S.S. \cite{Aharonian:2008aa,Aharonian:2009ah} 
and Fermi-LAT \cite{Abdo:2009zk} have invoked extensive discussions 
on the possible existence of dark matter (DM) signals (for a recent
review, see e.g., \cite{Bergstrom:2009ib,He:2009ra}). Meanwhile, the 
ratio of antiproton to proton measured by PAMELA \cite{Adriani:2008zq}
is well consistent with the astrophysical expectation from
interactions between CR nuclei and interstellar medium (ISM)
\cite{Donato:2008jk}. It indicates that if the DM
contributes to the electron/positron excesses, it should 
dominantly annihilate or decay into leptons instead of gauge
bosons or quarks \cite{Cirelli:2008pk,Yin:2008bs}. Although the
observational data of the electron spectra from ATIC and
Fermi-LAT/H.E.S.S. are not fully consistent, it has been shown in
literature that the DM models with annihilation or decay modes
directly to leptons can give good description to the observational
data, given proper mass of DM particle and flavors of the final
state particles (e.g.,
\cite{Cholis:2008wq,Nardi:2008ix,Bergstrom:2009fa,Meade:2009iu}).
Specifically, the ATIC data favor a $e^+e^-$ channel to describe
the bump and fast drop for energy $\sim 600$ GeV
\cite{Chang:2008zzr,Zhang:2008tb}. While for Fermi-LAT and
H.E.S.S. data, a softer electron/positron spectrum from the decay
of $\mu^+\mu^-$ or $\tau^+\tau^-$ final states can better
reproduce the smooth behavior
\cite{Bergstrom:2009fa,Meade:2009iu}.

However, for the numerical studies of fitting to the data in the
literature so far one usually takes some specific parameters for a
given model, then fits to the data  for illustrations of how the
model works instead of performing a global analysis. Therefore
bias exists in the conclusions drawn from this kind of studies. A
global fit to the observational data thus will be very useful to
extract the model-independent implication from the data and to
explore the correlations among different parameters. In this work
we employ a Markov Chain Monte Carlo (MCMC) technique
\cite{MCMC97,MacKayBook,Neil93} to globally fit the parameters of
the DM scenario to the experimental data on the electron/positron
spectrum observed. The constraint from the diffuse $\gamma$-ray 
emission around the Galactic center (GC), known as GC ridge, by 
H.E.S.S. \cite{Aharonian:2006au} is also included in the MCMC 
program. Based on the global fitting results, we can then further 
investigate the possible implication on the models of both on the 
particle physical side and also the astrophysical side of DM, in 
a manner of self-consistency.

Another issue which is not seriously taken into account
in previous studies is the influence of electron/positron background
on the DM models. The global fit method makes it possible to include
the uncertainties of the background contribution. This point is also
discussed in this work.

The paper is organized as follows. In Sec. II we introduce the
production mechanism of the electrons and positrons generated from
DM annihilation or decay, and their propagation effect in the
Milky Way (MW). In Sec. III we describe the accompanied
$\gamma$-rays which are served as a cross check of the
self-consistency of the model configuration. The MCMC fitting
procedure and results are given in Sec. IV. Finally, Sec. V is the
summary.

\section{Production and propagation of the cosmic $e^{\pm}$}

Due to the constraints from $\bar{p}/p$ by PAMELA, the DM is
thought to be dominately coupled with leptons
\cite{Cirelli:2008pk,Yin:2008bs}. Since in this work we mainly
focus on the methodology of using MCMC to globally fit the
observational data, we will limit our discussion in the cases that
DM annihilates or decays to lepton pairs directly\footnote{A more
general discussion including the quarks or gauge bosons in the
final states will be a straightforward extension of the current
study and will be present in a future publication.}. The
production rate of electrons (positrons) can be written as
\begin{equation}
q_e({\bf r},E)=\frac{\langle\sigma v\rangle}{2m_{\chi}^2}\left.\frac{{\rm d}N}
{{\rm d}E}\right|_i\rho^2({\bf r}),
\label{annisource}
\end{equation}
for DM annihilation, or
\begin{equation}
q_e({\bf r},E)=\frac{1}{m_{\chi}\tau}\left.\frac{{\rm d}N}
{{\rm d}E}\right|_i\rho({\bf r}),
\label{decaysource}
\end{equation}
for DM decay, where $m_\chi$ is the mass of DM particle, $\langle\sigma
v\rangle$ or $\tau$ are the annihilation cross section or decay age of
DM respectively, $\left.\frac{{\rm d}N}{{\rm d}E}\right|_i$ is the
electron yield spectrum for one annihilation or decay with $i=e,\,\mu,\,
\tau$, and $\rho({\bf r})$ is DM spatial density in the MW halo.
The density profile of the MW halo is taken as the form
\begin{equation}
\rho(r)=\frac{\rho_{s}}{(r/r_s)^{\gamma}(1+r/r_s)^{3-\gamma}},
\label{profile}
\end{equation}
where $\gamma$ represents the central cusp slope of the density
profile, $r_s$ and $\rho_s$ are scale radius and density
respectively. For the MW DM halo, we adopt the total mass to be
$M_{\rm MW} \approx 10^{12}$ M$_{\odot}$ \cite{Xue:2008se} and the
concentration parameter to be $c_{\rm MW}\approx 13.5$
\cite{Bullock:1999he}. Then we have $r_s=r_{\rm MW}/c_{\rm
MW}(2-\gamma)$ where $r_{\rm MW}\approx 260$ kpc is the virial
radius of MW halo. Then $\rho_s$ can be derived by requiring
$M_{\rm MW}=\int \rho {\rm d}V$. The local density $\rho_\odot$
in this process is checked to be within $0.27$ to $0.25$ GeV cm$^{-3}$
for $\gamma$ varying from $0$ to $1.5$.

After the production from DM annihilation or decay, the electrons
and positrons will propagate diffusively in the MW due to the
scattering with the random magnetic field. Besides the diffusion,
the dominant process of electron propagation is the energy loss
from synchrotron radiation in Galactic magnetic field and the
inverse Compton (IC) scatterings in the interstellar radiation
field (ISRF). There may also be global convection driven by
stellar wind and reacceleration by random interstellar shock,
however, as shown in Ref. \cite{Delahaye:2008ua}, these effects
can be safely neglected for electron energy $\gtrsim 10$ GeV. The
propagation equation is
\begin{equation}
-D\nabla^2 N({\bf r},E) + \frac{\partial}{\partial E}\left[\frac{{\rm d}E}
{{\rm d}t}N({\bf r},E)\right]=q({\bf r},E),
\end{equation}
where $D(E)=\beta D_0{\cal R}^{\delta}$ (${\cal R}=pc/Ze$ is the
rigidity) is the diffusion coefficient, ${\rm d}E/{\rm
d}t=-\epsilon^2/\tau_{\rm E}$ with $\epsilon=E/1{\rm\ GeV}$ and
$\tau_{\rm E}\approx 10^{16}$ s, is the energy loss rate for
typical values of the Galactic magnetic field and ISRF, $q({\bf
r},E)$ is the source function of $e^{\pm}$ as given in Eqs.
(\ref{annisource}) and (\ref{decaysource}), and $N({\bf r},E)$ is
the propagated $e^{\pm}$ spectrum.

The propagator for a point source located at $(r,z)$ from the solar
location with monochromatic injection energy $E_{\rm S}$ can be written
as \cite{Lavalle:2006vb,Lavalle:1900wn}
\begin{equation}
{\cal G}_{\odot}(r,z,E\leftarrow E_{\rm S})=\frac{\tau_{\rm E}}
{E\epsilon}\times \hat{\cal G}_{\odot}(r,z,\hat{\tau}),
\label{propposi}
\end{equation}
in which we define a pseudo time $\hat{\tau}$ as
\begin{equation}
\hat{\tau}=\tau_{\rm E}\frac{\epsilon^{\delta-1}-
\epsilon_{\rm S}^{\delta-1}}{1-\delta}.
\end{equation}
$\hat{\cal G}_{\odot}(r,z,\hat{\tau})$ is the Green's function for
the re-arranged diffusion equation with respect to the pseudo time
$\hat{\tau}$
\begin{equation}
\hat{\cal G}_{\odot}(r,z,\hat{\tau})=\frac{\theta(\hat{\tau})}
{4\pi D_0\hat{\tau}}\exp\left(-\frac{r^2}{4D_0\hat{\tau}}\right)
\times {\cal G}^{\rm 1D}(z,\hat{\tau}).
\end{equation}
The effect of boundaries along $z=\pm L$ appears in ${\cal G}^{\rm 1D}$
only. Following Ref. \cite{Lavalle:2006vb} we use two distinct regimes to
approach ${\cal G}^{\rm 1D}$:
\begin{itemize}

\item for $\zeta\equiv L^2/4D_0\hat{\tau}\gg 1$ (the extension of electron
sphere $\lambda\equiv\sqrt{4D_0\hat{\tau}}$ is small)
\begin{equation}
{\cal G}^{\rm 1D}(z,\hat{\tau})=\sum_{n=-\infty}^{\infty}(-1)^n
\frac{\theta(\hat{\tau})}{\sqrt{4\pi D_0\hat{\tau}}}\exp\left(
-\frac{z_n^2}{4D_0 \hat{\tau}}\right),
\end{equation}
where $z_n=2Ln+(-1)^nz$;

\item otherwise 
\begin{equation}
{\cal G}^{\rm 1D}(z,\hat{\tau})=\frac{1}{L}\sum_{n=1}^{\infty}
\left[\exp(-D_0k_n'^2\hat{\tau})\phi_n'(0)\phi_n'(z)\right],
\end{equation}
where
\begin{eqnarray}
\phi_n(z)&=&\sin[k_n(L-|z|)];\ \ k_n=(n-1/2)\pi/L,\\
\phi_n'(z)&=&\sin[k_n'(L-z)];\ \ k_n'=n\pi/L.
\end{eqnarray}

\end{itemize}
For any source function $q(r,z,\theta;E_{\rm S})$ the local observed
flux of positrons can be written as
\begin{equation}
\Phi_{\odot}=\frac{v}{4\pi}\times 2\int_0^L{\rm d}z
\int_0^{R_{\rm max}}r{\rm d}r\int_E^{\infty}{\rm d}E_{\rm S}
{\cal G}_{\odot}(r,z,E\leftarrow E_{\rm S})
\int_0^{2\pi}{\rm d}\theta q(r,z,\theta;E_{\rm S}).
\label{fluxposi}
\end{equation}

For the propagation parameters, we use the medium (referred as ``MED'') set
of parameters which is derived through fitting the observational B/C data
given in Ref. \cite{Donato:2003xg}, i.e., $D_0=0.0112$ kpc$^{2}$~Myr$^{-1}$,
$\delta=0.70$ and the height of the diffusive halo $L=4$ kpc.

\section{Gamma rays}

For the purely leptonic annihilation or decay of DM particles,
$\gamma$-rays can be generally produced in two ways. One is the
final state radiation (FSR) which is emitted directly from the
external legs when DM particles annihilates or decays to charged
leptons\footnote{Note that for the tau channel, the decay of
$\tau$ leptons will produce a large number of  neutral pions which
can then decay into photons. We also include this contribution in
the FSR.}. The other is the IC scattering photons from ISRF when
the electrons and positrons propagate in the MW. Compared with the
IC photons, the FSR has several advantages. First the spectrum of
FSR is unique and may give smoking gun diagnostic for DM signal.
In addition, the FSR does not depend on the astrophysical
environment such as the distribution of ISRF. Therefore in this
work we only consider the FSR. It will simplify the calculation,
meanwhile the results obtained on the constraints on the model
parameters is also conservative.

The photon yield spectrum for $e^{\pm}$ or $\mu^{\pm}$ channel for
$m_{\chi}\gg m_e,\,m_{\mu}$ can be written as
\cite{Bergstrom:2004cy,Beacom:2004pe}
\begin{equation}
\left.\frac{{\rm d}N}{{\rm d}x}\right|_i=\frac{\alpha}{\pi}
\frac{1+(1-x)^2}{x}\log\left(\frac{s}{m_i^2}(1-x)\right),
\label{fsr_emu}
\end{equation}
where $\alpha\approx 1/137$ is the fine structure constant, $i=e,\,\mu$.
For DM annihilation we have $s=4m_{\chi}^2$ and $x=E_{\gamma}/m_{\chi}$;
while for DM decay $s=m_{\chi}^2$ and $x=2E_{\gamma}/m_{\chi}$
\cite{Essig:2009jx}. For $\tau^{\pm}$ channel, we adopt the total
parameterization including the direct FSR component as shown in
Eq.(\ref{fsr_emu}) and the decay products from the chain $\tau \rightarrow
\pi^0 \rightarrow \gamma$ \cite{Fornengo:2004kj}
\begin{equation}
\left.\frac{{\rm d}N}{{\rm d}x}\right|_{\tau}=x^{-1.31}\left(6.94x
-4.93x^2-0.51x^3\right)e^{-4.53x},
\label{fsr_tau}
\end{equation}
with the same definition of $x$ as above.

The $\gamma$-ray flux along a specific direction can be written as
\begin{eqnarray}
\phi(E_{\gamma},\psi)&=&C\times W(E_{\gamma})\times J(\psi) \nonumber\\
            &=&\left\{
            \begin{array}{ll}
            \frac{\rho_{\odot}^2R_{\odot}}{4\pi}\times\frac{\langle
            \sigma v\rangle}{2m_{\chi}^2}\frac{{\rm d}N}{{\rm d}E_{\gamma}}
            \times\frac{1}{\rho_{\odot}^2R_{\odot}}\int_{LOS}\rho^2(l)
            {\rm d}l,&{\rm for\ annihilation}\\
            \frac{\rho_{\odot}R_{\odot}}{4\pi}\times\frac{1}{m_{\chi}\tau}
            \frac{{\rm d}N}{{\rm d}E_{\gamma}}\times\frac{1}{\rho_{\odot}
            R_{\odot}}\int_{LOS}\rho(l){\rm d}l,&{\rm for\ decay}
            \end{array}\right.
\label{jpsi}
\end{eqnarray}
where the integral is taken along the line-of-sight (LOS), $W(E)$ and $J(\psi)$
represent the particle physics factor and the astrophysical factor
respectively, $R_{\odot}=8.5$ kpc is the solar system location from GC,
and $\rho_{\odot}$ is the local DM density. For the emission from a
diffuse region with solid angle $\Delta\Omega$, we define the average
astrophysical factor as
\begin{equation}
J_{\Delta\Omega}=\frac{1}{\Delta\Omega}\int_{\Delta\Omega}J(\psi){\rm
d}\Omega,
\end{equation}
which is fully determined by the parameter $\gamma$ in Eq.(\ref{profile}).

In this work we use the diffuse $\gamma$-ray observation in the
sky region $|l|<0.8^\circ$ and $|b|<0.3^\circ$ around the GC by
H.E.S.S. \cite{Aharonian:2006au} to constrain the central profile
of DM density distribution. It should be noted that the H.E.S.S. 
observation of the GC ridge $\gamma$-ray flux is a background 
subtracted one. In Ref. \cite{Aharonian:2006au} the emission from 
$0.8^\circ<|b|<1.5^\circ$, $|l|< 0.8^\circ$ is taken as the 
background. Therefore the H.E.S.S. reported result is not the 
total emission of the selected sky region, but a lowered one.
To compare the calculated result with the data, we calculate 
the $\gamma$-ray flux from DM in the H.E.S.S. signal region 
($|b|<0.3^\circ$, $|l|< 0.8^\circ$) with subtracting the one in 
the H.E.S.S. background region ($0.8^\circ<|b|<1.5^\circ$, 
$|l|< 0.8^\circ$). Thus the $J_{\Delta\Omega}$ factor adopted for 
the $\gamma$-ray calculation actually means $J_{\Delta\Omega}^{\rm sig}
-J_{\Delta\Omega}^{\rm bkg}$ in the following.

Similar to Ref. \cite{Bi:2009am}, we employ a power-law component 
$\phi = a_{\gamma}E_{\gamma}^{-b_{\gamma}}$ to represent the 
astrophysical background contribution to the $\gamma$-rays. The FSR from 
DM is then added to the background to fit the observational data. The 
fit to the $\gamma$-ray data is also included in the MCMC code so that 
we can derive a consistent constraint on the DM density profile.


\section{MCMC Method and results}
\label{Method}
\subsection{Method}

In our study, we perform a global analysis employing the MCMC
technique to determine the parameters related to DM models. The
MCMC sampler is implemented by using the Metropolis-Hastings
algorithm. It is an efficient procedure for generating samples
which lies in the fact that its equilibrium distribution
corresponds to the likelihood function in parameter space.
Comparing with a traditional approach where the likelihood
function is evaluated on a grid of points in parameter space, the
MCMC method has the great potential in expanding the dimension of
the parameter series, since the computational requirements of MCMC
procedures are insensitive to the dimensionality of the parameter
space.

Fixing the $e^+e^-$ backgrond, we have the following parameter  space:
\begin{equation}
\label{parameter} {\bf P} \equiv (m_{\chi}, \langle\sigma v\rangle
{\ \rm or\ \tau},B_{e}, B_{\mu}, B_{\tau}, \gamma, a_\gamma, b_\gamma),
\end{equation}
where $m_{\chi}$ is the DM particle mass, $\langle\sigma v\rangle$
is the annihilation cross section, for decaying DM $\tau$ denotes
the lifetime,  $B_{e}$, $B_{\mu}$ and $B_{\tau}$ are the
annihilation (or decay)  branching ratios of DM particle into
$e^+e^-$, $\mu^+\mu^-$ and $\tau^+\tau^-$ pairs. For the scenario
considered in this work, we have the constraint condition
$B_e+B_{\mu}+B_{\tau}\equiv 1$. The parameter $\gamma$
parametrizes the DM density profile and $a_\gamma$, $b_\gamma$ 
are parameters characterizing astrophysical $\gamma$-ray background.
So in our global fitting procedure $7$ free parameters are in 
general involved.

With variables in (\ref{parameter}) we can calculate the 
theoretical expectations of electrons and positrons produced from DM. 
However, to compare with the observational data, we still need to give 
the astrophysical background contribution to the electrons and positrons. 
Since the background calculation, especially for positrons, is quite 
complicated and consumes a lot of time, we employ two approaches
of the background. Firstly we fix the background calculated by GALPROP
package \cite{galprop} based on a conventional diffusion + convection 
(DC) model \cite{Lionetto:2005jd,Yin:2008bs}. It is shown that the
conventional GALPROP model can give fairly good description to the 
CR data before PAMELA/ATIC/Fermi-LAT/H.E.S.S. \cite{Yin:2008bs}, as 
well as the all-sky diffuse $\gamma$-ray data from EGRET (except the 
``GeV excess'', \cite{Strong:1998fr}) and Fermi-LAT at intermediate 
latitudes \cite{Abdo:2009ka}. The GALPROP model parameters are the 
same as Ref. \cite{Yin:2008bs} except for the high energy electron 
injection spectrum we adopt a slightly softer one $2.58$. For the
second approach we include a more complete treatment on this problem.
A power law function $q_{e^-}=a_{e^-}E_{e^-}^{-b_{e^-}}$ with two
free parameters $a_{e^-},\,b_{e^-}$ is involved to describe the
injection source of primary electrons. For background positrons 
which are thought to be produced through CRs interact with ISM, 
we use the locally measured proton and Helium spectra as parameterized 
in Ref. \cite{Donato:2008jk} and an average ISM density $\sim 1$ 
cm$^{-3}$ to calculate the positron production rate. For the secondaries
production from $p-p$ inelastic collision we use the parameterization 
given in Ref. \cite{Kamae:2006bf}. The interaction
is restricted in a thin disk with half height $\sim 0.1$ kpc. The
spatial distribution of positron source is neglected, which is 
demonstrated not to affect the final positron spectrum 
\cite{Delahaye:2008ua}. In addition, we employ a free factor $c_{e^+}$
lies between $0.5$ and $2$ to describe the possible uncertainties
about the ISM density and the disk height. That is to say, totaly
we will have $10$ free parameters in this approach.

\subsection{Data sets}

Since the electron spectrum measurements between ATIC and
Fermi-LAT/H.E.S.S. still have discrepancy, for an unbiased study
at the current stage we will try to fit two different data sets
respectively. One is to combine the PAMELA positron fraction data,
ATIC electron data\footnote{The ATIC data include ATIC1+ATIC2+ATIC4, 
which are taken from the talk by J. Isbert in {\it TANGO in PARIS} 
workshop, see http://irfu.cea.fr/Meetings/tangoinparis/main.htm. 
From the numerical calculation, we find that the fitting results 
between ATIC1+ATIC2+ATIC4 and the published ATIC1+ATIC2 data are 
very similar, but the former one converges much faster and better.}
and the H.E.S.S. $\gamma$-ray data (Data set
I), and the other is the combination of PAMELA, Fermi-LAT, H.E.S.S.
electron and H.E.S.S. $\gamma$-ray data (Data set II). Note that
for PAMELA we use only the data with energies higher than $\sim 5$
GeV in the MCMC fitting code. The low energy data are thought to
be more easily to be affected by the solar modulation. We get the
$\chi^2$ by comparing the spectrum between the theoretical
expectations and the corresponding observational values. For the
PAMELA, the observation quantity is the positron fraction
$e^{+}/(e^{+}+e^{-})$, while for ATIC, Fermi-LAT and H.E.S.S. the
undistinguished fluxes of positrons and electrons are concerned.
We have taken the total likelihood to be the products of the
separate likelihoods ${\bf \cal{L}}$ of all the data. In other
words, defining $\chi^2_i \equiv -2 \log {\bf {\cal{L}}_{i}}$, we
get
\begin{equation}
\label{chi2}
\chi^2_{\rm total} = \sum_{i} \chi^2_{i},
\end{equation}
where $i$ labels different observational data.


\subsection{Numerical Results}

\subsubsection{Fixed $e^+e^-$ background}

\begin{figure}[!htb]
\begin{center}\vspace{10mm}
\includegraphics[width=0.4\columnwidth]{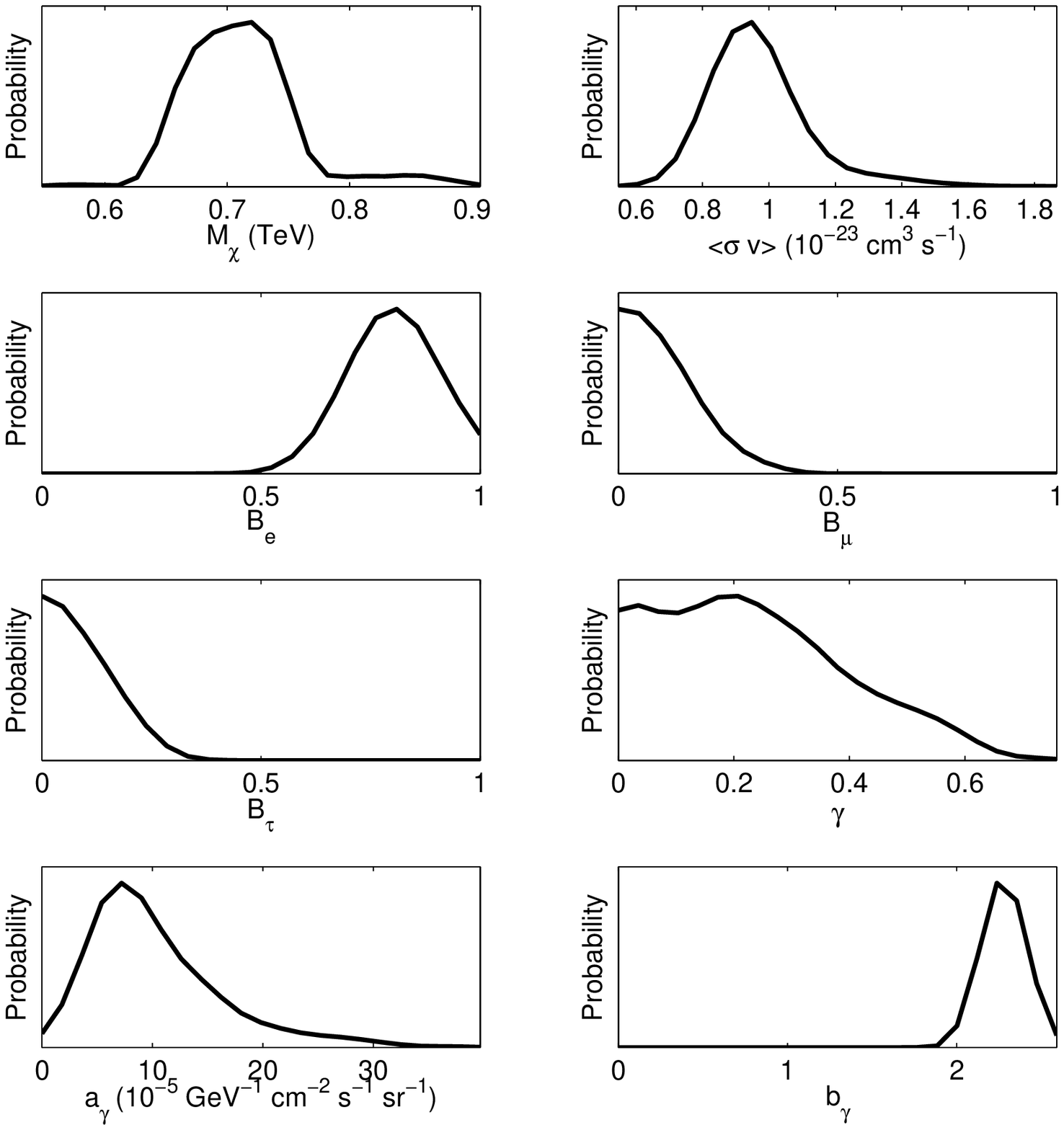}
\hspace{10mm}
\includegraphics[width=0.4\columnwidth]{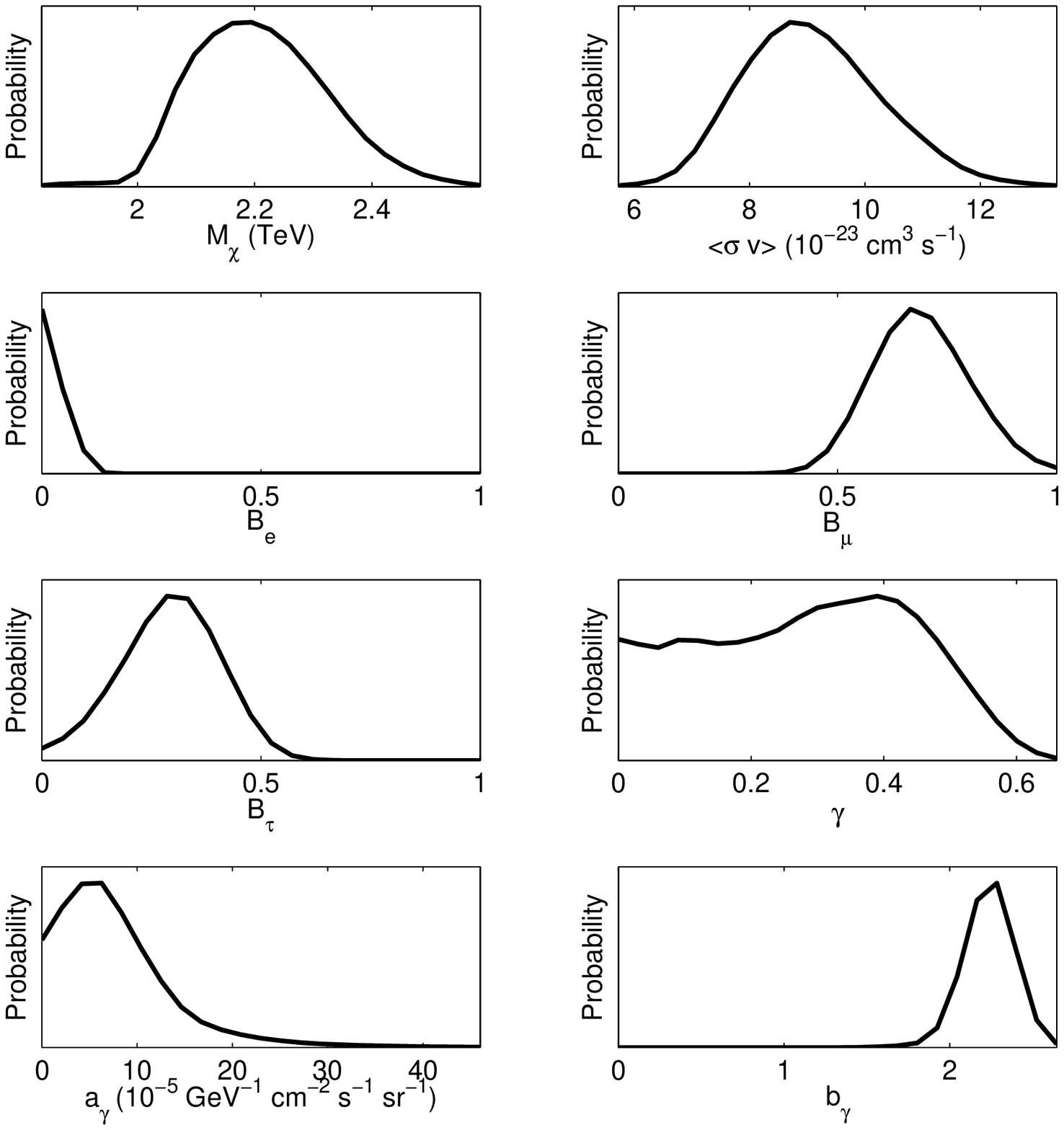}
\caption{Probability distributions of the eight parameters in annihilation
DM scenario used to fit Data set I ({\it left}) and II ({\it right})
respectively, for fixed $e^+e^-$ approach. To get the $1D$ probability 
of each parameter we have marginalized over the other parameters.
\label{fig:anni}}
\end{center}
\end{figure}

Firstly we consider the fixed background approach. The one dimensional 
probability distributions of the fitting parameters for annihilation 
DM scenario are shown in Fig.\ref{fig:anni}. The {\it left} eight 
panels are for Data set I, and the {\it right} eight panels are for 
Data set II respectively. To get the $1D$ probability of each parameter 
we have marginalized over the other parameters. The details about
the parameters are compiled in Table \ref{table:anni}.

To reproduce the peak of ATIC observation of Data set I, the mass
of DM is found to be $\sim 0.70$ TeV with annihilation channel
almost purely to $e^+e^-$. The $2\sigma$ upper limits for
$\mu^+\mu^-$  and $\tau^+\tau^-$ channels are $B_{\mu}<0.269$ and 
$B_{\tau}<0.226$ respectively. While for the fit to Data set II, a
larger mass $m_{\chi}\approx 2.2$ TeV with larger $B_{\mu}$ and
$B_{\tau}$ which can give softer electron/positron spectra is
favored. The $e^+e^-$ channel which can give strongly peaked
electron/positron spectra is remarkably suppressed. The $2\sigma$
upper limit for $B_{e}$ is $\sim 0.03$.

An important issue is the constraint from GC diffuse $\gamma$-rays
observed by H.E.S.S.. A strong constraint on the central DM
density profile is shown. At $2\sigma$ level, we find
$\gamma<0.539$ for Data set I and $<0.515$ for Data set II. It is
significantly flatter than the canonical NFW profile with
$\gamma=1$ \cite{Navarro:1996gj}. This indicates that DM
particle might not be so cold, but behaves like warm DM instead
\cite{robert,Bi:2009am}. We can also note that the probability 
distributions of $B_{\mu}$ and $B_{\tau}$ are relatively broad 
for the Data set II fit. The reason for this is due to the 
degeneracy between $\mu^+\mu^-$ and $\tau^+\tau^-$ channels. 
We will discuss this issue later.

\begin{figure}[!htb]
\begin{center}\vspace{10mm}
\includegraphics[width=0.4\columnwidth]{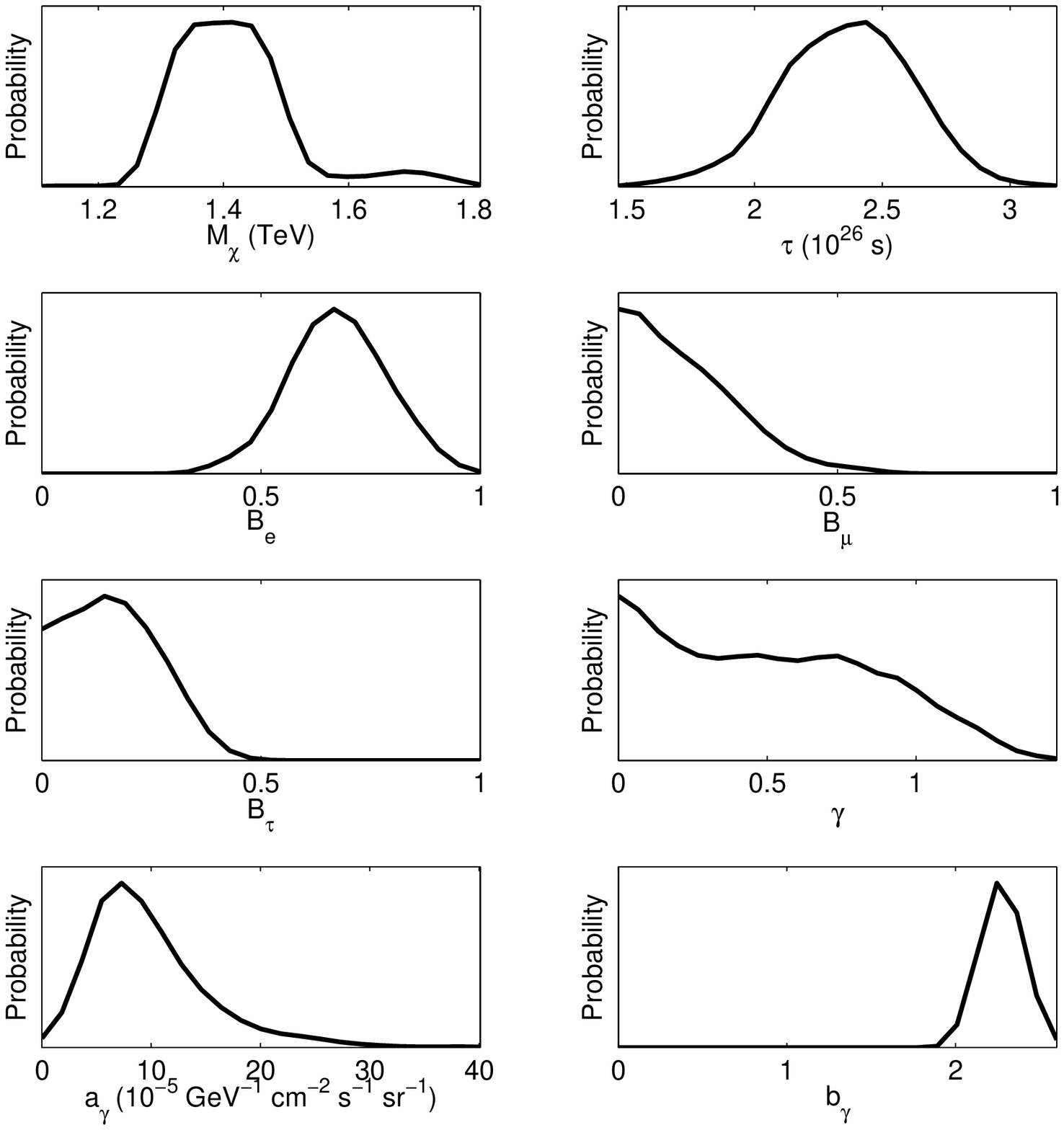}
\hspace{10mm}
\includegraphics[width=0.4\columnwidth]{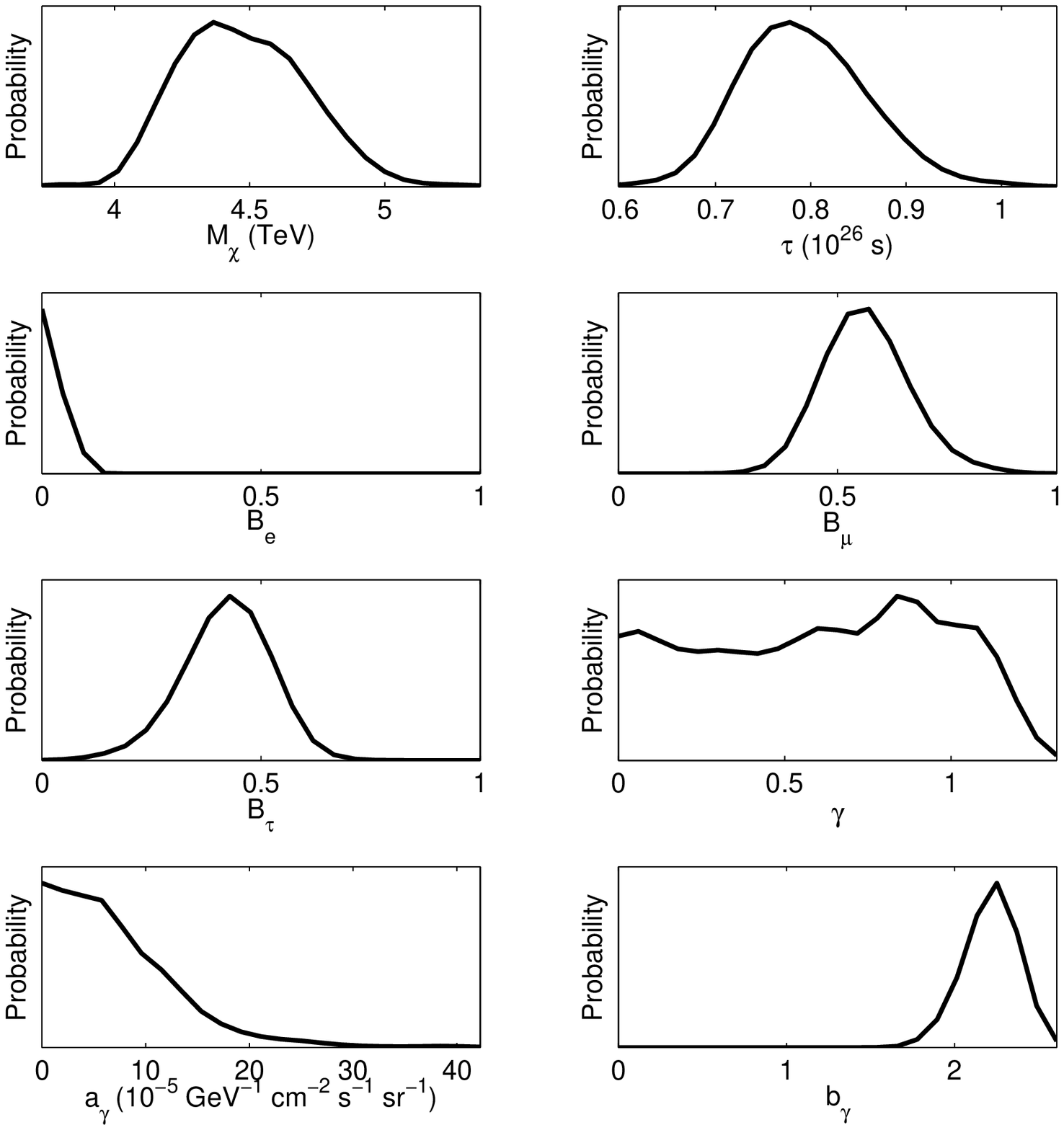}
\caption{The same as Fig.\ref{fig:anni} but for decaying DM scenario.
\label{fig:decay}}
\end{center}
\end{figure}

\begin{table}[!htb]
\centering
\caption{Mean $1~\sigma$ errors or $95\%$ limits for the parameters in annihilation DM 
scenario for the fixed $e^+e^-$ background approach. 
}
\begin{tabular}{|c|c|c|}
\hline 
Parameters & Data Set I & Data Set II\\
\hline
$m_\chi$ (TeV)& $0.711_{-0.042}^{+0.032}$ & $2.211_{-0.111}^{+0.112}$\\
\hline
$\langle\sigma v\rangle$ ($10^{-23}$ cm$^3$ s$^{-1}$) & $0.969_{-0.129}^{+0.115}$&$9.105_{-1.172}^{+1.177}$\\
\hline
$B_e$ &$0.795\pm0.090$& $<0.031$\\
\hline
$B_{\mu}$&$<0.269$ & $0.691_{-0.050}^{+0.043}$\\
\hline
$B_{\tau}$& $<0.226$&$0.293_{-0.044}^{+0.050}$\\
\hline
$\gamma$& $<0.539$& $<0.527$\\
\hline
$a_\gamma$ ($10^{-5}$ GeV$^{-1}$ cm$^{-2}$ s$^{-1}$ sr$^{-1}$) &$10.441_{-3.479}^{1.197}$& $<19.608$\\
\hline
$b_\gamma$ & $2.279_{-0.033}^{+0.031}$& $2.236_{-0.035}^{+0.048}$\\
\hline
\end{tabular}
\label{table:anni}
\end{table}

\begin{table}[!htb]
\centering
\caption{Mean $1~\sigma$ errors or  $95\%$ limits for the parameters
in decaying DM 
scenario for the fixed $e^+e^-$ background approach.}
\begin{tabular}{|c|c|c|}
\hline 
Parameters & Data Set I & Data Set II\\
\hline
 $m_\chi$ (TeV)& $1.418_{-0.088}^{+0.064}$ & $4.475_{-0.233}^{+0.231}$\\
 \hline
 $\tau$ ($10^{26}$ s) & $2.359_{-0.246}^{+0.252}$ &$0.794_{-0.063}^{+0.066}$\\
 \hline
 $B_e$ &$ 0.675_{-0.050}^{+0.049}$ &$<0.025$\\
 \hline
 $B_{\mu}$ &$<0.387$ &$0.565_{-0.044}^{+0.032}$\\
 \hline
 $B_{\tau}$ & $<0.333$ &$0.421_{-0.033}^{+0.046}$\\
 \hline
 $\gamma$ &  $<1.114$ &  $<1.130$\\
 \hline
 $a_\gamma$ ($10^{-5}$ GeV$^{-1}$ cm$^{-2}$ s$^{-1}$ sr$^{-1}$) & $9.936_{-3.102}^{+1.130}$ &  $<18.498$\\
 \hline
 $b_\gamma$& $2.275_{-0.035}^{+0.030}$& $2.215_{-0.033}^{+0.066}$\\
\hline
\end{tabular}
\label{table:decay}
\end{table}

The results for decaying DM model are shown in Fig. \ref{fig:decay}
and Table \ref{table:decay} respectively. We can see that the
results are very similar to the annihilation case except for the
constraint on the density profile parameter $\gamma$. It is just
as expected that since the electron/positron source function is
proportional to $\rho$ for DM decay instead of the $\rho^2$
dependence of DM annihilation, the constraint on the central cusp
slope is expected to be much weaker.

\begin{figure}[!htb]
\begin{center}\vspace{10mm}
\includegraphics[width=0.45\columnwidth]{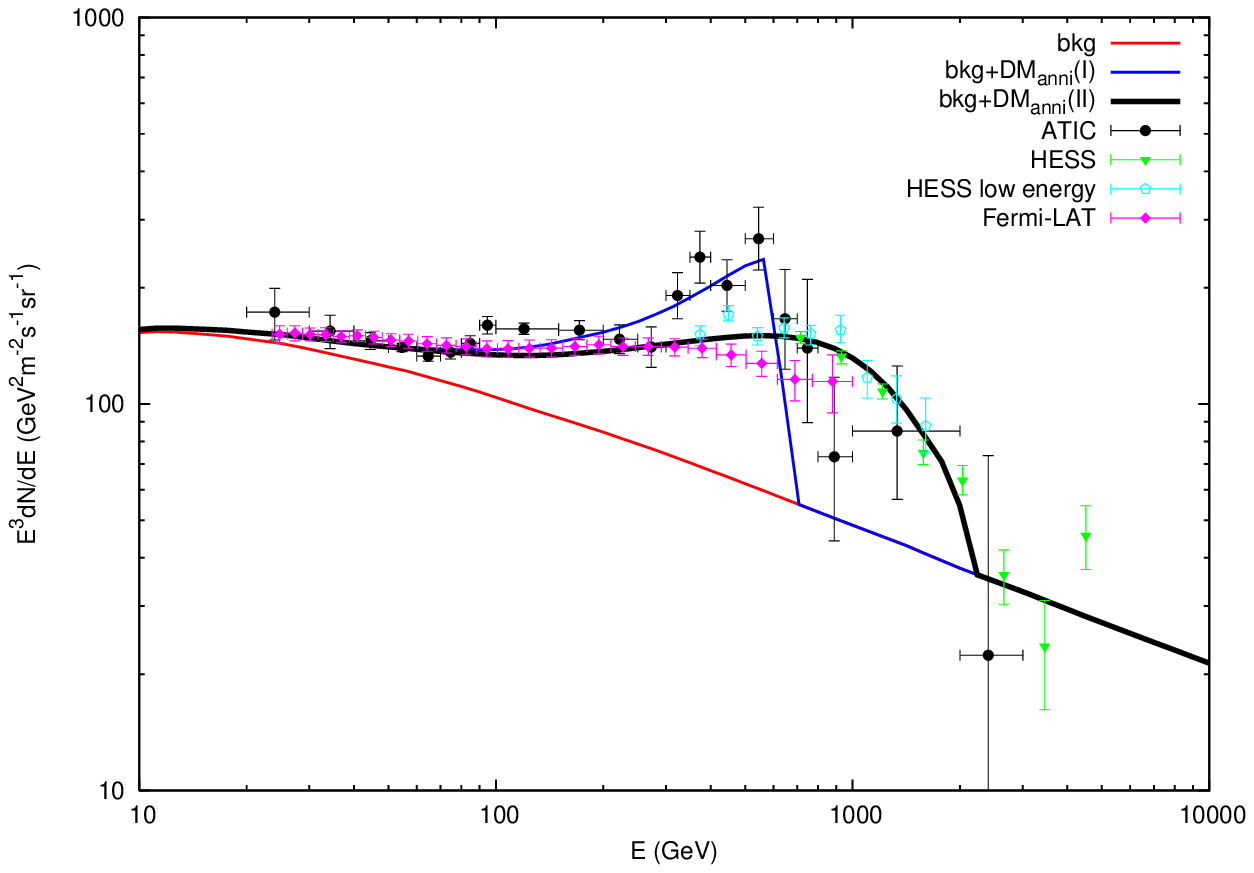}
\includegraphics[width=0.45\columnwidth]{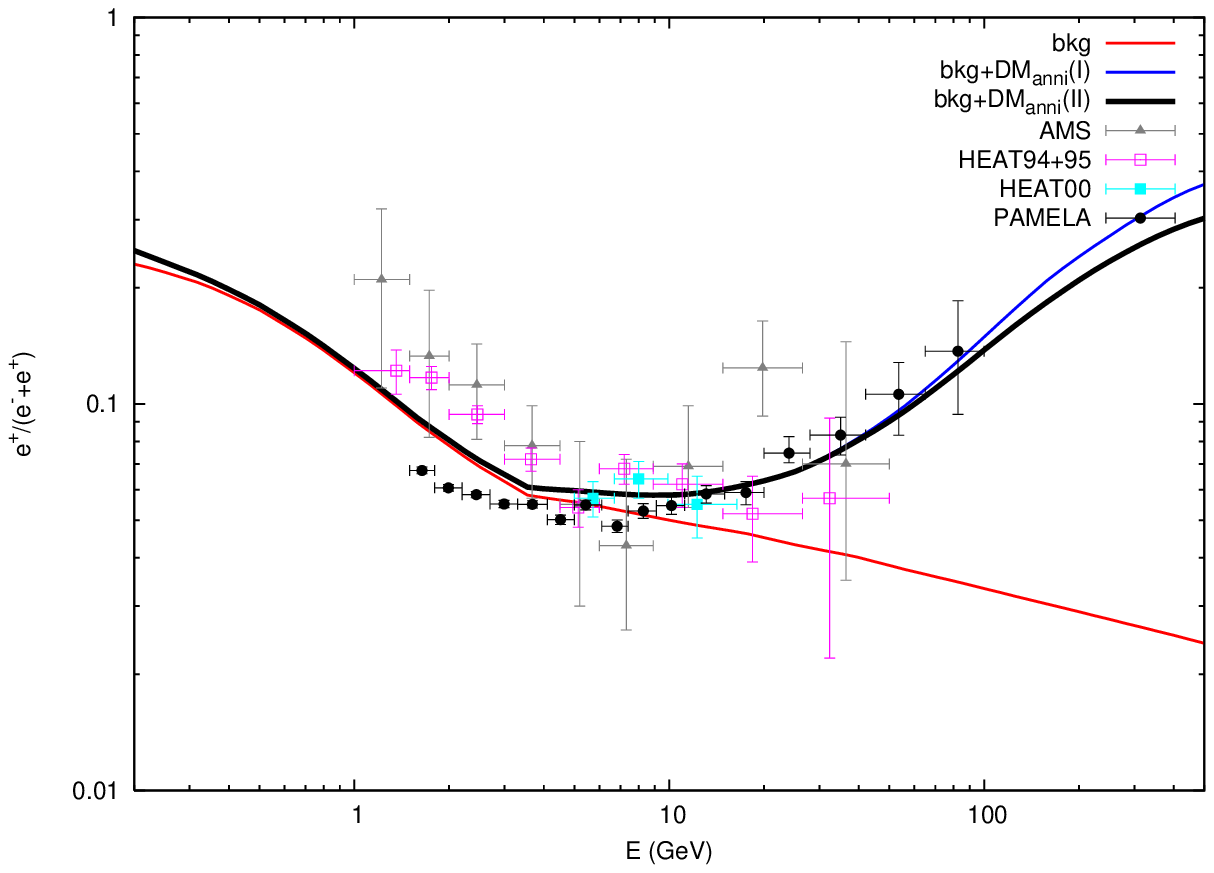}
\caption{{\it Left:} the $e^++e^-$ spectrum including the contribution
from DM annihilation compared with the observational data from ATIC
\cite{Chang:2008zzr}, HESS \cite{Aharonian:2008aa,Aharonian:2009ah} and
Fermi-LAT \cite{Abdo:2009zk}. {\it Right:} the $e^+/(e^-+e^+)$ ratio
including the contribution from DM annihilation as a function of energy
compared with the data from AMS \cite{Aguilar:2007yf},
HEAT \cite{Barwick:1997ig,Coutu:2001jy} and PAMELA \cite{Adriani:2008zr}.
\label{ep:anni}}
\end{center}
\end{figure}

In Fig. \ref{ep:anni} we show the results of the total 
electron + positron spectra and positron fraction for
the best-fitting parameters, for annihilation DM scenario. The 
results show a very good agreement with the observational data. 
The results for decaying DM scenario, which are not shown here, 
are almost undistinguishable from the annihilation case.
Because the final states from DM annihilation and decay
are completely the same, the only difference comes from the
spatial distribution of DM induced electrons/positrons. 
While the observed high energy electrons/positrons should mainly come from
the local regions near the Earth where there is no big difference
between the $\rho^2$ distribution for DM annihilation and the
$\rho$ distribution for DM decay, therefore the propagated
electron/positron spectra are very similar for both.

\begin{figure}[!htb]
\begin{center}\vspace{10mm}
\includegraphics[width=0.45\columnwidth]{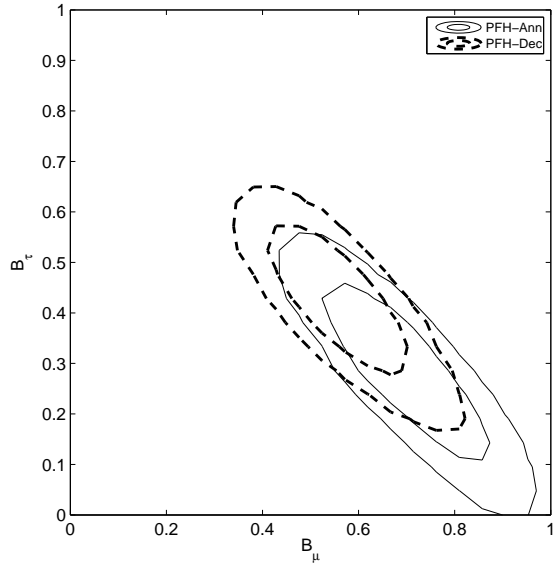}
\caption{Two dimensional plot for the correlation between
parameters $B_{\mu}$ and $B_{\tau}$ for the fit to Data set II,
for DM annihilation (solid) and decay (dashed) respectively. The
inner and outer contours show the results at $1\sigma$ and
$2\sigma$ confidence levels. \label{fig:bmutau}}
\end{center}
\end{figure}

Finally we investigate the correlation among the various
annihilation or decay channels. We find that there is no
evident correlation between parameter $B_e$ and other two
branching ratios. This can be understood that the spectrum of 
electrons/positrons from $e^+e^-$
channel has a very hard and spiky signature, which can be easily
distinguished from the other two channels. However, the cases will
be different for $\mu^+\mu^-$ and $\tau^+\tau^-$ channels. For the
$\mu^+\mu^-$ or $\tau^+\tau^-$ channels, the electrons are
generated through the decay of muons or tauons, which both can
give a softer electron spectrum. Thus the results from
$\mu^+\mu^-$ channel and $\tau^+\tau^-$ channel should have some
degeneracies. In Fig. \ref{fig:bmutau} we plot the 2D correlation
between $B_{\mu}$ and $B_{\tau}$ for the fit to Data set II, for 
DM annihilation and decay respectively. It indeed shows a strong 
anti-correlation between $B_{\mu}$ and $B_{\tau}$. That means 
to describe the Data set II the muon channel is equivalent to the 
tauon channel to some extent. It will be not easy to distinguish 
whether the DM particle couples mainly
with muon flavor or tauon flavor through the electron/positron
data. This degeneracy will also lead to an uncertainty on the study
of $\gamma$-ray prediction or constraint. For tauon channel, more
$\gamma$-ray photons can be generated through tauon decay, while
for muon channel the FSR radiated $\gamma$-ray photons will be
much fewer. On the other hand, we expect future accurate $\gamma$-ray
experiment to break this degeneracy.

\subsubsection{Varying $e^+e^-$ background}

\begin{figure}[!htb]
\begin{center}\vspace{10mm}
\includegraphics[width=0.4\columnwidth]{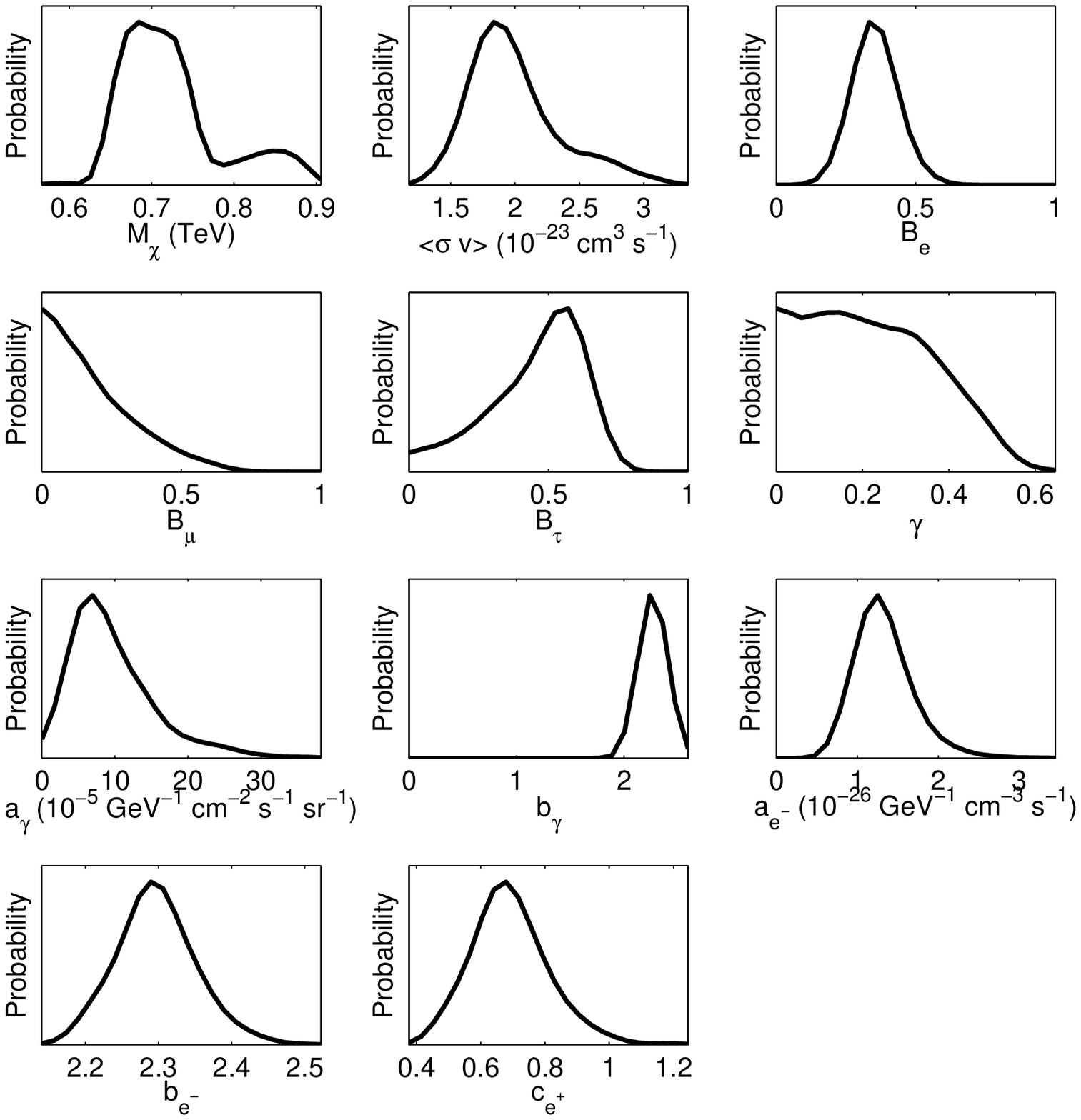}
\hspace{10mm}
\includegraphics[width=0.4\columnwidth]{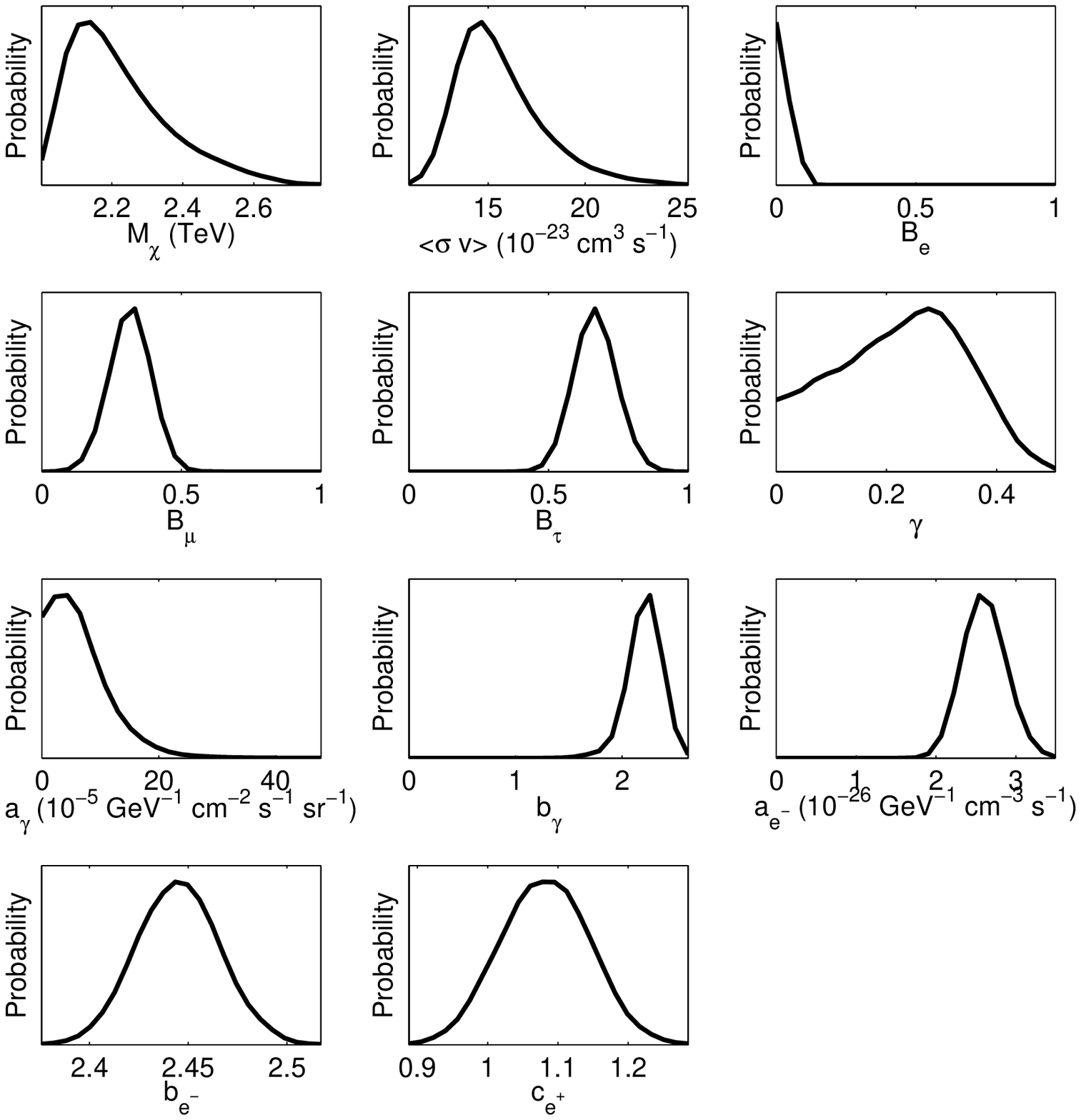}
\caption{Probability distributions of the eleven parameters in annihilation
DM scenario used to fit Data set I ({\it left}) and II ({\it right})
respectively, for varying $e^+e^-$ background approach. 
\label{fig:bkganni}}
\end{center}
\end{figure}

\begin{figure}[!htb]
\begin{center}\vspace{10mm}
\includegraphics[width=0.4\columnwidth]{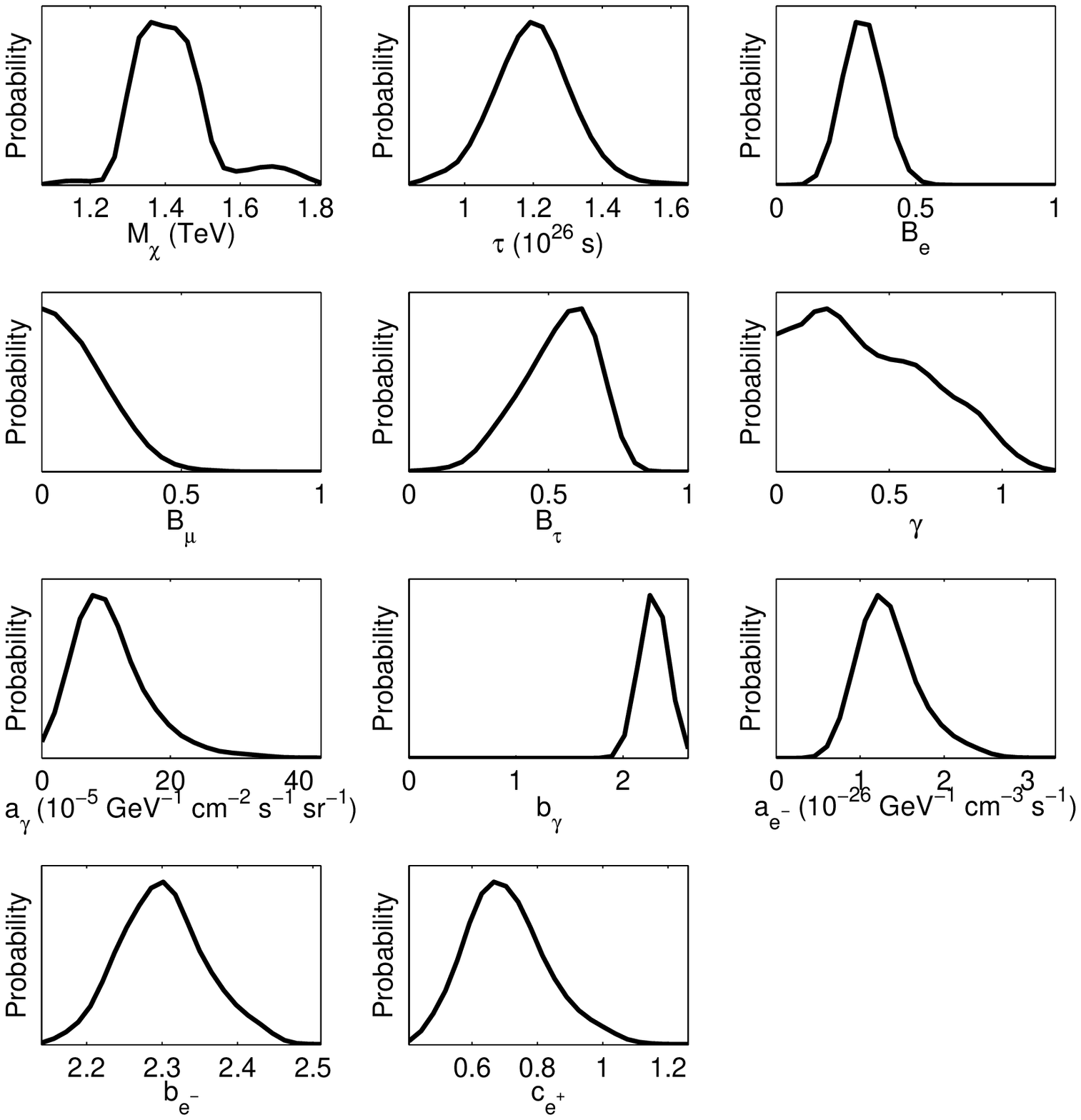}
\hspace{10mm}
\includegraphics[width=0.4\columnwidth]{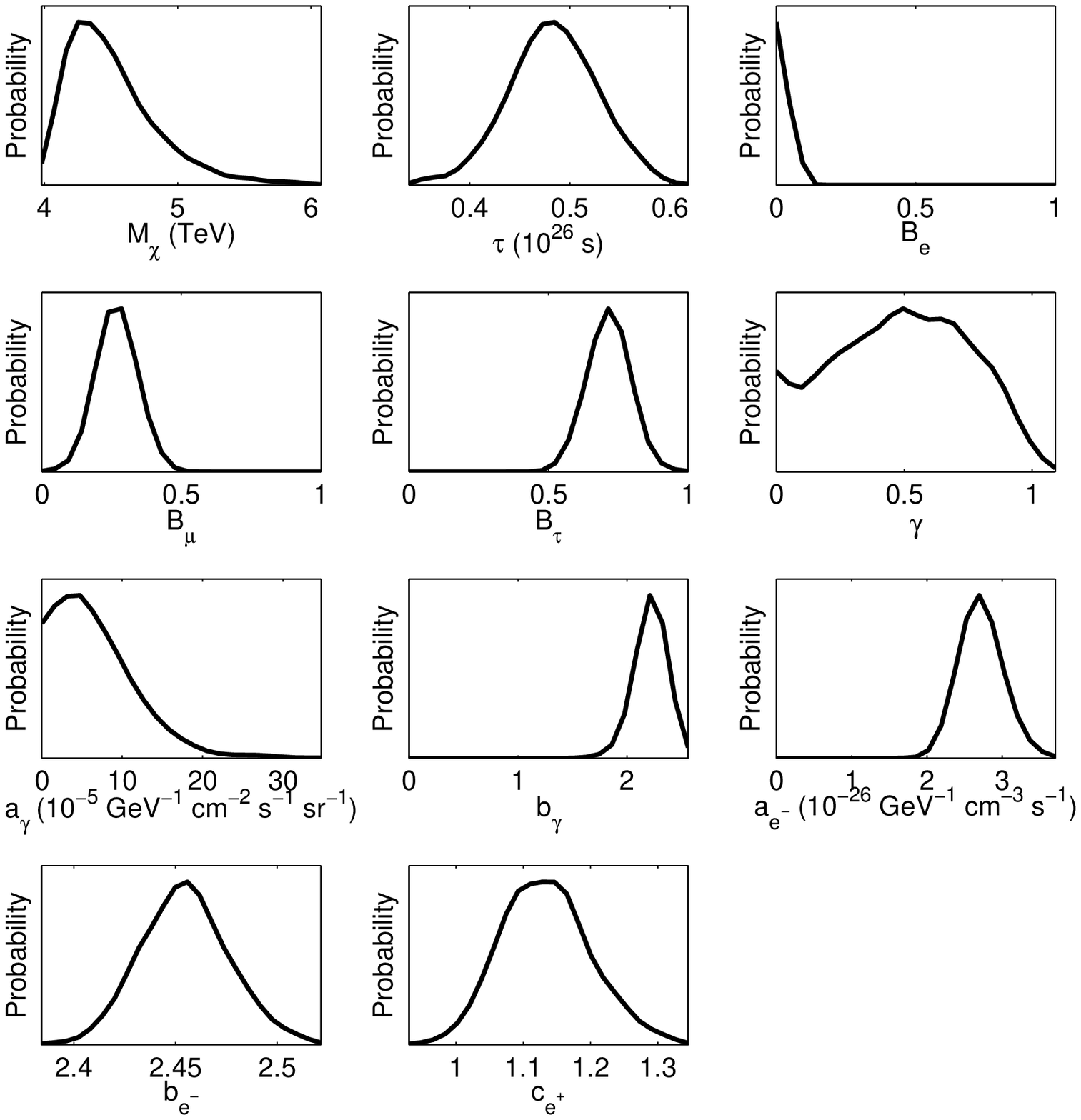}
\caption{The same as Fig.\ref{fig:bkganni} but for decaying DM scenario.
\label{fig:bkgdecay}}
\end{center}
\end{figure}

In this section, we present the results by setting the $e^+e^-$ 
background free. Three additional parameters $a_{e^-}$, $b_{e^-}$ 
and $c_{e^+}$ are added in the MCMC approach. The $1D$ parameter
distributions are shown in Figs. \ref{fig:bkganni}, \ref{fig:bkgdecay}
and Tables \ref{table:bkganni}, \ref{table:bkgdecay} respectively,
for the four cases as considered in the fixed $e^+e^-$ background
section. 

We find that there are remarkable effects on the branching ratios 
if changing the background. This is reasonable since a different
background will lead to a different blank for DM to fill. For example
if the background electron spectrum is harder, then the required
contribution from DM is expected to be softer. Actually if we relax
the background electron spectrum, a harder spectrum compared with the 
previous fixed one is favored according to the MCMC fit, for both
data sets. Thus the branching ratio to $\tau^+\tau^-$ is expected 
to be larger, while $B_e$ is suppressed. This is extremely evident 
for Data set I, where $\sim 1/3$ to $e^+e^-$ channel is enough to fit 
the data and the best-fit $B_\tau$ becomes $>0.5$ ($0.639$ for 
annihilation and $0.655$ for decay modes). Since in this case 
the best-fit background spectrum is about $3.15$ after propagation, 
compared with the one $\sim 3.35$ of the fixed background, 
the required electron contribution from DM is much softer. However,
a non-negligible branching ratio to $e^+e^-$ is still needed to
describe the peak of ATIC data. A combination of $\mu^+\mu^-$ and
$\tau^+\tau^-$ is also favored for Data set II, but with $\tau^+\tau^-$
channel dominating instead, compared with the fixed background 
approach. For other parameters the quanlitative constraints are almost
unchanged for different background choices.

The changes of the fitting branching ratios between different
background choices suggest that we should be cautious when claiming 
the DM properties used to explain the data, given the uncertain 
knowledge about the background. While global fitting will be the 
only way to seperate the background and possible ``signal'', and 
derive model-independent implications of the data.

\begin{table}[!htb]
\centering
\caption{Mean $1~\sigma$ errors or  $95\%$ limits for the parameters 
in annihilation DM scenario, for the varying $e^+e^-$ background approach. 
}
\begin{tabular}{|c|c|c|}
\hline 
Parameters & Data Set I & Data Set II\\
\hline
$m_\chi$ (TeV)& $0.723_{-0.055}^{+0.063}$ & $2.221_{-0.184}^{+0.141}$\\
\hline
$\langle\sigma v\rangle$ ($10^{-23}$ cm$^3$ s$^{-1}$) & $1.996_{-0.311}^{+0.341}$ & $15.625_{-1.972}^{+2.072}$\\
\hline
$B_e$ &$0.355_{-0.034}^{+0.029}$& $<0.017$\\
\hline
$B_{\mu}$&$<0.486$ & $0.316_{-0.025}^{+0.027}$\\
\hline
$B_{\tau}$& $0.459_{-0.052}^{+0.099}$ & $0.667_{-0.029}^{+0.026}$\\
\hline
$\gamma$& $<0.472$& $<0.402$\\
\hline
$a_\gamma$ ($10^{-5}$ GeV$^{-1}$ cm$^{-2}$ s$^{-1}$ sr$^{-1}$) & $9.793_{-3.328}^{+1.264}$ & $<16.565$\\
\hline
$b_\gamma$ & $2.270_{-0.034}^{+0.036}$& $2.211_{-0.033}^{+0.057}$\\
\hline
$a_e^-$ ($10^{-26}$ GeV$^{-1}$ cm$^{-3}$ s$^{-1}$) & $1.330_{-0.176}^{+0.095}$ & $2.609_{-0.116}^{+0.101}$\\
\hline
$b_e^-$ & $2.297_{-0.053}^{+0.052}$ & $2.445\pm0.020$\\
\hline
$c_e^+$ & $0.691_{-0.111}^{+0.112}$& $1.082\pm0.063$\\
\hline
\end{tabular}
\label{table:bkganni}
\end{table}

\begin{table}[!htb]
\centering
\caption{Mean $1~\sigma$ errors or $95\%$ limits for the parameters
in decaying DM  scenario for the varying $e^+e^-$ background approach.}
\begin{tabular}{|c|c|c|}
\hline 
Parameters & Data Set I & Data Set II\\
\hline
 $m_\chi$ (TeV)& $1.420_{-0.094}^{+0.067}$ & $4.501_{-0.308}^{+0.305}$\\
 \hline
 $\tau$ ($10^{26}$ s) & $1.198_{-0.106}^{+0.107}$ &$0.483_{-0.042}^{+0.043}$\\
 \hline
 $B_e$ &$0.312_{-0.030}^{+0.026}$ &$<0.026$\\
 \hline
 $B_{\mu}$ &$<0.366$ &$0.269_{-0.030}^{+0.027}$\\
 \hline
 $B_{\tau}$ & $0.534_{-0.051}^{+0.078}$ & $0.715_{-0.030}^{+0.031}$\\
 \hline
 $\gamma$ &  $<0.917$ &  $<0.893$\\
 \hline
 $a_\gamma$ ($10^{-5}$ GeV$^{-1}$ cm$^{-2}$ s$^{-1}$ sr$^{-1}$) & $10.924_{-3.170}^{+1.306}$ &  $<16.377$\\
 \hline
 $b_\gamma$& $2.287_{-0.030}^{+0.033}$& $2.216_{-0.034}^{+0.057}$\\
\hline
$a_e^-$ ($10^{-26}$ GeV$^{-1}$ cm$^{-3}$ s$^{-1}$) & $1.352_{-0.196}^{+0.093}$ & $2.720_{-0.129}^{+0.100}$\\
\hline
$b_e^-$ & $2.301_{-0.056}^{+0.057}$& $2.455\pm0.021$\\
\hline
$c_e^+$ & $0.708_{-0.117}^{+0.120}$& $1.134\pm0.065$\\
\hline
\end{tabular}
\label{table:bkgdecay}
\end{table}

With including additional parameters, we find that the constraints on each 
parameter are broader comparing with the fixed backgrond approach, however,
the goodness of fit defined as $G.O.F=\chi^2/d.o.f$ is getting better (Tab \ref{tab:gof}).

\begin{table}
\centering
\caption{The reduced $\chi^2_r$ of each case.}
\begin{tabular}{|c|c|c|c|c|}
\hline
&\multicolumn{2}{c|}{Fixed}&\multicolumn{2}{c|}{Varying}\\
\hline
&~Anni~&~Decay~&~Anni~&~Decay~\\
\hline
Data Set I&1.439&1.434&1.065&1.075\\
\hline
Data Set II&1.143&1.151&1.057&1.112\\
\hline
\end{tabular}
\label{tab:gof}
\end{table}


\section{Summary}
\label{Sum}

In this work we employ a MCMC technique to determine the parameters
of DM scenario in interpreting the recent observations of electron
spectra and positron fraction data. Both the DM annihilation and
decay are considered in this work. The DM particle is  assumed to
couple only with leptons. Considering the discrepancy between
experimental data from ATIC and Fermi-LAT/H.E.S.S., we fit two data
sets independently. We find that for Data set I (PAMELA+ATIC), DM
with $m_{\chi}\approx0.7$ TeV for annihilation (or $1.4$ TeV for
decay) and a non-negligible $e^+e^-$ component is favored. For Data 
set II (PAMELA+Fermi-LAT+H.E.S.S.) $m_{\chi}\approx 2.2$ TeV for 
annihilation (or $4.5$ TeV for decay) and the combination of 
$\mu^+\mu^-$ and $\tau^+\tau^-$ final states can best fit the data. 
There is degeneracy between parameters $B_{\mu}$ and $B_{\tau}$. The
constraint on the central density profile of DM halo from the
diffuse $\gamma$-rays near the GC is also included in the MCMC code
self-consistently. We find the H.E.S.S. observations of the GC
$\gamma$-rays can give strong constraint on the DM density slope of
the central cusp for the annihilation DM scenario. In that case the
NFW profile with $\gamma=1$, which is regarded as the typical
predication from cold DM scenario, is excluded with a high
significance ($>3\sigma$). For the decaying DM scenario, the
constraint is much weaker. 

We find that the largest uncertainties on the determination of 
DM branching ratios come from the limited understanding of the $e^+e^-$
background. Besides using the canonical background expectations from
CR propagation models, we further consider the approach to involve
the background uncertainties. Three additional free parameters are
adopted to describe the background contribution and fitted globally
with the DM parameters. It is shown that the branching ratios depend
sensitively on the background results. The limited knowledge about
the background actually prevents us from a precise determination
of the DM final state information. However, with the accumulation
of more observational data from PAMELA, AMS02 and many other upcoming
experiments with much higher precision and detailed information of
each species, we may expect to get better understanding on both the
background and the ``exotic signal'' (if exists). While global 
fitting will be the only way to separate the background and signal, 
and derive model-independent implications of the data. Since the 
MCMC method works in a full high-dimensional parameter space, the 
code developed in this work will be a useful tool in studying both
the CR physics and the exotic new physics, given more precise 
observational data in the future.

We do not include the radio constraints on the synchrotron radiation
of the models in this work, though the radio observations in the GC 
region might set even stronger constraints than $\gamma$-rays, as 
studied in Refs. \cite{Regis:2008ij,Bertone:2008xr,Bergstrom:2008ag}.
The radio constraints are found to be sensitive to the DM profile,
instead of the central most configuration of the magnetic field. 
However, since the knowledge of the magnetic field relies on some 
assumptions such as the accretion state of central black hole and 
the equipartition of the matter kinetic energy with the magnetic
pressure \cite{Bertone:2008xr}, we think there are still uncertainties 
of the radio constraints. Additionally, one expects that antiproton
constraints would be important for the tau channel, which has not 
yet been considered in the current work. This combined with the radio 
bound may provide a stringent constraint for the $\tau$ channel, which 
deserves a detailed study in future. 

As an end, we would like to mention that, there exists a self-consistent 
way to explain the current data by introducing a light mediator state in 
DM annihilation (or decay) \cite{ArkaniHamed:2008qn}, which has 
not yet been considered in the current work. The distinct advantages of 
this type of models are that, the strongly constrained decay into hadrons 
is kinematically forbidden, and Sommerfeld enhancement with small velocity
of Galactic DM allows for the large annihilation cross sections required 
by observations. This kind of model is not included in this work due to 
the fact that, we need to introduce more parameters in the fitting
such as the mass of the mediator which can not be effectively constrained
by the current data on the one hand, and the present work is the first 
step of the application of MCMC method in DM indirect searches on the
other hand. We hope the forthcoming improvement of the method can 
partially address this issue.


\section*{Acknowledgements}

We thank Jun-Qing Xia, Yi-Fu Cai for discussion. This work is supported 
in part by National Natural Science Foundation of China under Grant 
Nos. 90303004, 10533010, 10675136 and 10803001 and by the Chinese Academy of 
Science under Grant No. KJCX3-SYW-N2.


\end{document}